\documentclass[a4paper,12pt]{article}
\usepackage{amsmath}
\usepackage{amssymb}
\usepackage{amsfonts}
\usepackage{a4wide}
\usepackage{txfonts}
\usepackage{indentfirst}
\usepackage{times}

\newcommand {\oks}[2]{{\raise0.7ex\hbox{${\scriptstyle #1}$}\!\mathord{\left/
{\vphantom{{1}{2}}}\right.\kern-\nulldelimiterspace}\!\lower0.7ex
\hbox{${\scriptstyle #2}$}}}

\begin{document}

\title{
\bf Radiative transitions of high energy neutrino \\ in dense
matter}

\author{\bf A. E. Lobanov\thanks{E-mail: lobanov@th466.phys.msu.ru}}
\date{}
\maketitle
\begin{center}
{\em Moscow State University, Department of Theoretical physics,
$119992$  Moscow, Russia}
\end{center}

\begin{abstract}
{The quantum theory of the ``spin light'' (electromagnetic radiation
emitted  by a massive neutrino propagating in dense matter due to
the weak interaction of a neutrino with background fermions) is
developed. In contrast to the Cherenkov radiation, this effect
does not disappear even if  the medium refractive index is assumed
to be equal to unity. The formulas for the transition rate and the
total radiation power are obtained. It is found out that radiation
of photons is possible only when the sign of the particle helicity
is opposite to that of the effective potential describing the
interaction of a neutrino (antineutrino) with the background
medium. Due to the radiative self-polarization the radiating
particle can change its helicity. As a result, the active
left-handed polarized neutrino (right-handed polarized
antineutrino) converting to the state with inverse helicity can
become practically ``sterile''. Since the sign of the effective
potential depends on the neutrino flavor and the matter structure,
the ``spin light'' can change a ratio of active neutrinos of
different flavors. In the ultra relativistic approach, the
radiated photons averaged energy  is equal to one third of the
initial neutrino energy, and two thirds of the energy are carried
out by the final ``sterile'' neutrinos. This fact can be important
for the understanding of the ``dark matter'' formation mechanism
on the early stages of evolution of the Universe.}
\end{abstract}

A massive neutrino propagating in  dense matter can emit
electromagnetic radiation due to the weak interaction of a
neutrino with background fermions \cite{L49,L55}. As a result of
the radiation, neutrino can change its helicity due to the
radiative self-polarization. In contrast to the Cherenkov
radiation, this effect does not disappear even if the refractive
index of the medium is assumed to be equal to unity. This
conclusion is valid for any model of neutrino interactions
breaking spatial parity. The phenomenon was called the neutrino
``spin light'' in analogy with the effect, related with the
synchrotron radiation power depending  on the electron spin
orientation (see \cite{BTB95}).

The properties of the ``spin light'' were investigated basing upon the
quasi-classical theory of radiation and self-polarization of
neutral particles \cite{L38,L51} with the use of the
Bargmann--Michel--Telegdi equation \cite{BMT} and its
generalizations \cite{L43,L32}. This theory is valid when the
radiated photon energy is small as compared with the neutrino
energy, and this narrows the range of  astrophysical applications
of the obtained formulas.

In the present paper the properties of the ``spin light'' are
investigated  basing upon the consistent quantum theory, and this
allows the neutrino recoil in the act of radiation to be
considered for. The above mentioned restriction is eliminated in
this way. On the other hand, the detailed analysis of the results
of our investigations shows that the features of the effect depend
on the neutrino flavor, helicity and the matter structure. This
fact leads to conclusion that the ``spin light'' can initiate
transformation of a neutrino from the active state to practically
``sterile'' state, and the inverse  process is also possible.

When the interaction of a neutrino with the background fermions is
considered to be coherent, the propagation of a massive neutrino
in the matter is described by the Dirac equation with the
effective potential \cite{Wolf,MS}. In what follows, we restrict
our consideration to the case of a homogeneous and isotropic
medium. Then in the frameworks of the minimally extended standard
model, the form of this equation is uniquely determined by the
assumptions similar to those adopted in \cite{F}:
\begin{equation}\label{1}
  \left(i
  \hat{\partial}-\frac{1}{2}\hat{f}(1+\gamma^{5})-m_{\nu}\right)
  \varPsi_{\nu}=0.
\end{equation}
\noindent The function $f^{\mu}$ is a linear combination of
fermion currents and polarizations, its coefficients can depend on
these vectors squared. If the medium is at rest and unpolarized
then ${\bf f} =0.$ The component $f^{0}$ calculated in the first
order of the perturbation theory is as follows \cite{PP,N,NR}:
\begin{equation}\label{2}
f^{0}=\sqrt{2}G_{\mathrm F}\bigg\{\sum\limits_{f}^{}
\left(I_{e\nu} + T_{3}^{(f)}-2Q^{(f)}\sin^{2}\theta_{\mathrm
W}\right)(n_{f}-n_{\bar{f}})\bigg\}.
\end{equation}
\noindent Here, $n_{f},n_{\bar{f}}$ are the number densities of
background fermions and anti-fermions, $Q^{(f)}$ is the electric
charge of the fermion and $T_{3}^{(f)}$ is the third component of
the weak isospin for the left-chiral projection of it. The
parameter \mbox{$I_{e\nu}=1$} is equal to unity for the
interaction of electron neutrino with electrons. In other cases
$I_{e\nu}= 0.$ Summation is performed over all fermions $f$ of the
background.

Let us consider the process of emitting photons by a massive
neutrino in  unpolarized matter at rest. The formula for the
spontaneous radiation transition probability of a neutral
fer\-mion with anomalous magnetic moment $\mu_{0}$ is{\footnote{In
the expression for the radiation energy $\mathcal E,$ the
additional factor $k$ -- the energy of radiated photon -- appears
in the integrand.}}:
\begin{equation}\label{ar1}
\begin{array}{c}
  \displaystyle P=-\frac{1}{2p^{0}}\!\int\!\! d^{4}x\,d^{4}y\!\int
  \frac{d^{4}q\,d^{4}k}{(2\pi)^{6}}\,
  \delta(k^{2})\delta(q^{2}\!-m^{2}_{\nu})\times
  \\[8pt]
\displaystyle \times {\mathrm{Sp}}
\big\{\varGamma_{\mu}(x)\varrho_{i}(x,y;p,\zeta_{i})
\varGamma_{\nu}(y)\varrho_{f}(y,x;q,\zeta_{f})\big\}
\varrho^{\mu\nu}_{ph}(x,y;k).
\end{array}
\end{equation}
\noindent \noindent Here,
$\varrho_{i}(x,y;p),\;\varrho_{f}(y,x;q)$ are density matrices of
initial $(i)$ and final $(f)$ states of the fermion,
$\,\varrho^{\mu\nu}_{ph}(x,y;k)$ is the radiated photon density
matrix, $\varGamma^{\mu} = -\,\sqrt{4\pi}
\mu_{0}\sigma^{\mu\nu}k_{\nu}$ is the vertex function. The density
matrix of longitudinally polarized neutrino in the unpolarized
matter at rest constructed with the use of the solutions of
equation (\ref{1}) has the form
\begin{equation}\label{arx4}
\varrho(x,y;p,\zeta)=\frac{1}{2}\Delta_{p\zeta}^{2}(\hat{p}+m_{\nu})
(1-\zeta\gamma^{5}\hat{S}_{
p})e^{-i(x^{0}-y^{0})(p^{0}+f^{0}/2)+i({\bf x}-{\bf y}) {\bf
p}\Delta_{p\zeta}},
\end{equation}
\noindent where $p^{\mu}$ is the neutrino kinetic moment,
$\Delta_{p \zeta} = 1+\zeta f^{0}/2|{\bf{p}}|,$ and $S_{p}^{\mu}=
\left\{|{\bf p}|/m_{\nu},
 p^{0}{\bf p}/|{\bf p}|m_{\nu}\right\}.$
\noindent Thus, $ \zeta = \pm 1$ correspond to the sign of the
spin projection on the neutrino kinetic moment.

It is convenient to express the results of calculations using
dimensionless variables ${\gamma=p^{0}/m_{\nu}}$,
 $ d=|f^{0}|/2m_{\nu}$, 
  $\bar{\zeta}_{i,f}={\zeta}_{i,f}\,{\mathrm{sign}}(f^{0}).$
The transition rate under investi\-ga\-tion is defined as
\begin{equation}\label{q18}
\begin{array}{c}
\displaystyle W_{\bar{\zeta}_{f}}=\frac{\mu_{0}^{2}m_{\nu}^{3}}{4}
\Big\{(1+ \bar{\zeta}_{f})\left[Z(z_{1},1)
  \Theta(\gamma -\gamma_{1})
  +Z(z_{2},-1)\Theta(\gamma -\gamma_{2})\right]+\\[12pt]
\displaystyle +(1-\bar{\zeta}_{f})\left[Z(z_{1},1)
  \Theta(\gamma_{1} -\gamma)
  +Z(z_{2},-1)\Theta(\gamma_{2} -\gamma)\right]
  \Theta(\gamma -\gamma_{0})\Big\}(1-\bar{\zeta}_{i}).
\end{array}
\end{equation}
\noindent Here
\begin{equation}\label{q17}
\begin{array}{c}
\displaystyle
Z(z,\bar{\zeta}_{f})=\frac{1}{\gamma(\gamma^{2}-1)}\left\{\ln
z\left[\gamma^{2}
+d\sqrt{\gamma^{2}-1}+d^{2}+{1}/{2}\right]\right.+\\[12pt]
\displaystyle +\frac{1}{4}\left(z^{2}-z^{-2}\right)
\left[d^{2}\left(2\gamma^{2}-1\right)
+d\sqrt{\gamma^{2}-1}+{1}/{2}\right]+\\[12pt]
\displaystyle +\frac{\bar{\zeta}_{f}}{4}\left(z -z^{-1}\right)^{2}
\left[2d\sqrt{\gamma^{2}-1}+1\right]d\gamma-\\[12pt]
\displaystyle -\left(z  -z^{-1}\right)
\left[d^{2}  + d\sqrt{\gamma^{2}-1}+1\right]\gamma-\\[12pt]
\displaystyle -\left.\bar{\zeta}_{f}\left(z+z^{-1} -2\right)
\left[d\sqrt{\gamma^{2}-1}+\gamma^{2}\right]d\right\},
\end{array}
\end{equation}
where
$$ z_{1}=\displaystyle \gamma+\sqrt{\gamma^{2}-1}-2d,\;
z_{2}=\displaystyle \gamma-\sqrt{\gamma^{2}-1}+2d,$$ and
\begin{equation}\label{q15}
\begin{array}{ll}
\gamma_{0}=\displaystyle \sqrt{1+d^{2}}, & {}\\[8pt]
\displaystyle \gamma_{1}=\frac{1}{2}\left\{\left(1+2d\right)
 +\left(1+2d\right)^{-1}\right\},& {}\\[12pt]
 \displaystyle \gamma_{2}=\frac{1}{2}\left\{\left(1-2d\right)
 +\left(1-2d\right)^{-1}\right\},& d< 1/2,\\[16pt]
 \displaystyle \gamma_{2}= \infty, & d\geqslant 1/2.
\end{array}
\end{equation}
Therefore, the transition rate after summation  over polarizations
of the final neutrino is equal to
\begin{equation}\label{q19}
W_{\bar{\zeta}_{f}=1}+W_{\bar{\zeta}_{f}=-1}=
\frac{{\mu_{0}^{2}m_{\nu}^{3}}}{2}(1-\bar{\zeta}_{i})\Big\{Z(z_{1},1)
+Z(z_{2},-1)\Big\} \Theta(\gamma -\gamma_{0}).
\end{equation}

If $d\gamma \ll 1,$ then the expression (\ref{q18}) leads to the
formula
\begin{equation}\label{q20}
W_{\bar{\zeta}_{f}}=\frac{16\mu_{0}^{2}m_{\nu}^{3}d^{3}}{3\gamma}
(\gamma^{2}-1)^{3/2}(1-\bar{\zeta}_{i})(1+\bar{\zeta}_{f}),
\end{equation}
obtained in the quasi-classical approximation \cite{L55}. In the
ultra relativistic limit $(\gamma \gg 1, \; d\gamma \gg 1),$ the
transition rate is given by the expression
\begin{equation}\label{q21}
W_{\bar{\zeta}_{f}}=\mu_{0}^{2}m_{\nu}^{3}d^{2}{\gamma}
(1-\bar{\zeta}_{i})(1+\bar{\zeta}_{f}).
\end{equation}

Let us consider now the radiation power. If we introduce the
function
\begin{equation}\label{q22}
  \tilde{Z}(z,\bar{\zeta}_{f}) = \gamma Z(z,\bar{\zeta}_{f})
   - Y(z,\bar{\zeta}_{f}),
\end{equation}
\noindent where
\begin{equation}\label{q23}
\begin{array}{c}
\displaystyle
Y(z,\bar{\zeta}_{f})=\frac{1}{\gamma(\gamma^{2}-1)}\left\{-\ln
z\left[d^{2}
+d\sqrt{\gamma^{2}-1}+{1}\right]\gamma\right.-\\[12pt]
\displaystyle -\frac{1}{4}\left(z^{2}-z^{-2}\right) \left[d^{2}
+d\sqrt{\gamma^{2}-1}+{1}\right]\gamma+\\[12pt]
\displaystyle +\frac{1}{12}\left(z-z^{-1}\right)^{3}
\left[d^{2}\left(2\gamma^{2}-1\right)
+d\sqrt{\gamma^{2}-1}+{1}/{2}\right]+\\[12pt]
\displaystyle +
\frac{1}{2}\left(z-z^{-1}\right)\left[2d^{2}\gamma^{2}
+2d\sqrt{\gamma^{2}-1} + \gamma^{2} +{1}\right]+\\[12pt]
\displaystyle
+\frac{\bar{\zeta}_{f}}{12}\left(\left(z+z^{-1}\right)^{3}-8\right)
\left[2d\sqrt{\gamma^{2}-1}+1\right]d\gamma -\\[12pt]
\displaystyle -\left.\frac{\bar{\zeta}_{f}}{4}\left(z -z^{-1}
\right)^{2}\left[d\sqrt{\gamma^{2}-1}+\gamma^{2}\right]d \right\},
\end{array}
\end{equation}
\noindent then the formula for the total radiation power can be
obtained from (\ref{q18}), (\ref{q19}) by substitution
${Z}(z,\bar{\zeta}_{f}) \rightarrow \tilde{Z}(z,\bar{\zeta}_{f}).$
It can be verified that, if  $d\gamma \ll 1$ then the radiation
power is
\begin{equation}\label{q24}
I_{\bar{\zeta}_{f}}=\frac{32\mu_{0}^{2}m_{\nu}^{4}d^{4}}{3}
(\gamma^{2}-1)^{2}(1-\bar{\zeta}_{i})(1+\bar{\zeta}_{f}).
\end{equation}
\noindent This result was obtained in the quasi-classical
approximation \cite{L49}. In the ultra relativistic limit the
radiation power is equal to
\begin{equation}\label{q25}
I_{\bar{\zeta}_{f}}=\frac{1}{3}\mu_{0}^{2}m_{\nu}^{4}d^{2}{\gamma}^{2}
(1-\bar{\zeta}_{i})(1+\bar{\zeta}_{f}).
\end{equation}
\noindent It can be seen from equations (\ref{q21}) and
(\ref{q25}) that in the ultra relativistic limit the averaged
energy of emitted photons is $\langle \varepsilon_{\gamma} \rangle
= \varepsilon_{\nu}/3.$ It should be pointed out that the obtained
formulas are valid both for a neutrino and for an anti-neutrino.
The charge conjugation operation leads to the change of the sign
of the effective potential and the replacement of the left-hand
projector by the right-hand one in the equation (\ref{1}). Thus
the sign in front of the $\gamma^{5}$ matrix remains invariant.

The following conclusions can be deduced from the obtained
results. A neutrino (anti-neutrino) can emit photons due to
coherent interaction with  matter only  when its helicity has the
sign opposite to the sign of the effective potential. Otherwise,
radiation transitions are impossible. In the case of low energies
of the initial neutrino, only  radiation without spin-flip is
possible and the probability of the process is very small. At high
energies, the main contribution to radiation is given by
transitions with the spin-flip, the transitions without spin-flip
are either absent  or their probability is negligible. This
results in the effect of total self-polarization, i.~e. the
initially left-handed neutrino (right-handed anti-neutrino) are
transformed to practically ``sterile'' right-handed polarized
neutrino (left-handed polarized anti-neutrino). For ``sterile''
particles the situation is opposite.  They can be converted to the
active  form in the medium  ``transparent'' for the active
neutrino.

With the use of the effective potential calculated in the first
order of the perturbation theory (\ref{2}), the following
conclusions can be made. If the matter consists only of  electrons
then, in the framework of the minimally extended standard model in
the ultra relativistic limit (here we use gaussian units), we have
for the transition rate
\begin{equation}\label{q29}
\displaystyle
W_{\bar{\zeta}_{f}}=\frac{\alpha\varepsilon_{\nu}}{32\,\hbar}
\left(\frac{\mu_{0}}{\mu_{\mathrm B}}\right)^{2}
\left(\frac{\tilde{G}_{{\mathrm F}}\,n_{e}}{m_{e}c^{2}}\right)^{2}
(1-\bar{\zeta}_{i})(1+\bar{\zeta}_{f}),
\end{equation}
and for the total radiation power
\begin{equation}\label{q30}
\displaystyle
I_{\bar{\zeta}_{f}}=\frac{\alpha\varepsilon_{\nu}^{2}}{96\,\hbar}
\left(\frac{\mu_{0}}{\mu_{\mathrm B}}\right)^{2}
\left(\frac{\tilde{G}_{{\mathrm F}}\,n_{e}}{m_{e}c^{2}}\right)^{2}
 (1-\bar{\zeta}_{i})(1+\bar{\zeta}_{f}).
\end{equation}
\noindent Here $\varepsilon_{\nu}$ is the neutrino energy,
$\mu_{\mathrm B}=e/2m$ is the Bohr magneton, $\alpha $ is the fine
structure constant, $m_{e}$ is the electron mass and
$\tilde{G}_{{\mathrm F}}={G}_{{\mathrm
F}}(1+4\sin^{2}\theta_{\mathrm W}),$ where ${G}_{{\mathrm F}},\,
\theta_{\mathrm W}$ are the Fermi constant and the Weinberg angle
respectively. Thus, after the radiative transition, two thirds of
the initial active neutrino energy are carried out by the final
``sterile'' one.

At the same time, a muon neutrino in the electron medium does not
emit any radiation. Moreover, a muon neutrino does not emit
radiation in an electrically neutral medium, when the number
density of protons is equal to the electron number density. And an
electron neutrino can emit radiation if the electron number
density is greater than the neutron number density. An example of
such medium is the Sun. The neutron medium is ``transparent'' for
all active neutrinos, but an active antineutrino emits radiation
in such a medium,  the transition rate and the total radiation
power can be obtained from  equations (\ref{q29}) and (\ref{q30})
after substitution $\tilde{G}_{{\mathrm F}} \rightarrow
{G}_{{\mathrm F}}.$ Therefore the ``spin light'' can change the
ratio of active neutrino of different flavors.

It is obviously that the above conclusions change to opposite if
the matter consists of antiparticles. Therefore the neutrino
``spin light'' can serve as a tool for determination of the type
of astrophysical objects, since neutrino radiative transitions in
dense matter can result in radiating of photons of super-high
energies.  This effect can also be important  for the
understanding of the ``dark matter'' formation mechanism in early
stages of evolution of the Universe.

\bigskip

The author is very grateful to V.G.~Bagrov, A.V.~Borisov, and
V.Ch.~Zhukovsky for fruitful discussions.

\bigskip

This work was supported in part by the grant of President of
Russian Federation for leading scientific schools (Grant SS ---
2027.2003.2)

\end{document}